\documentclass{article}
\usepackage{spconf,amsmath,graphicx,hyperref, booktabs, comment}
\usepackage{siunitx}
\makeatletter
\def\title#1{\gdef\@title{\MakeUppercase{#1}}}
\def\name#1{\gdef\@name{\shortstack[c]{\itshape #1\\[0.5em]}}}
\makeatother

\title{A Point Process Model of Skin Conductance Responses in a Stroop Task \\ for Predicting Depression and Suicidal Ideation}
\name{Kleanthis Avramidis$^1$,\, Myzelle Hughes$^2$,\, Idan A Blank$^3$,\, Dani Byrd$^2$,\, Assal Habibi$^2$, \\ \em{Takfarinas Medani$^1$,\, Richard M Leahy$^1$,\, Shrikanth Narayanan$^{1,2}$}\thanks{This study was sponsored by the Defense Advanced Research Projects Agency (DARPA) under cooperative agreement No. N660012324006. The content of the information does not necessarily reflect the position or the policy of the Government, and no official endorsement should be inferred.}}
\address{$^1$ Viterbi School of Engineering, University of Southern California \\ $^2$ Dornsife College of Letters, Arts, and Sciences, University of Southern California \\ $^3$ Department of Psychology, University of California, Los Angeles}

\begin{document}
\ninept
\maketitle
\begin{abstract}
Accurate identification of mental health biomarkers can enable earlier detection and objective assessment of compromised mental well-being. In this study, we analyze electrodermal activity recorded during an Emotional Stroop task to capture sympathetic arousal dynamics associated with depression and suicidal ideation. We model the timing of skin conductance responses as a point process whose conditional intensity is modulated by task-based covariates, including stimulus valence, reaction time, and response accuracy. The resulting subject-specific parameter vector serves as input to a machine learning classifier for distinguishing individuals with and without depression. Our results show that the model parameters encode meaningful physiological differences associated  with depressive symptomatology and yield superior classification performance compared to conventional feature extraction methods.
\end{abstract}
\begin{keywords}
Electrodermal Activity, Point Process, Skin Conductance Response, Depression, Suicidal Ideation
\vspace{-0.2cm}
\end{keywords}

\section{Introduction}
\vspace{-0.2cm}
Depression is a major mental health disorder, affecting an estimated 280 million people worldwide~\cite{IHME_GBD_2023} with severe psychosocial and financial consequences. Its impact extends beyond mood disturbance, impairing cognitive function, decision-making, and attentional control~\cite{lawlor2020dissecting, wang2020attention}. Suicidal ideation is frequently comorbid with symptoms of major depression. Early and accurate identification of these mental health conditions is critical for improving treatment outcomes, yet current diagnostic practices rely heavily on self-reported symptoms and clinical interviews~\cite{bone2017signal}, both of which are susceptible to recall bias, subjectivity, and social stigma~\cite{baumeister2007psychology}. These challenges have driven the search for objective biomarkers that can complement existing clinical assessment methods.

Electrodermal activity (EDA) provides a non-invasive measure of physiological arousal elicited through the sympathetic nervous system. Rapid activations of skin conductance activity are typically associated with cognitive load, emotional reactivity, and/or attentional engagement~\cite{critchley2002electrodermal}. EDA is event-driven, hence sparse in information content, consisting of transient skin conductance responses (SCRs) triggered by discrete internal or external events. This temporal structure naturally motivates the use of point process modeling, which could represent the emergence of SCRs as a stochastic process modulated by experimental and behavioral covariates.

In the current study, we apply a point process generalized linear model (GLM) to EDA recorded from healthy, depressed, and suicidal individuals during an Emotional Stroop task. The model is designed to capture the conditional intensity of SCR occurrences as a function of trial-level features such as stimulus valence, reaction time, and response accuracy. The resulting subject-specific parameter estimates index the individual patterns of sympathetic arousal dynamics. We subsequently use those features for machine learning classification of individuals with depression and/or suicidal ideation. Our contributions are summarized as follows:

 \begin{figure}
    \centering
    \includegraphics[width=1.01\linewidth]{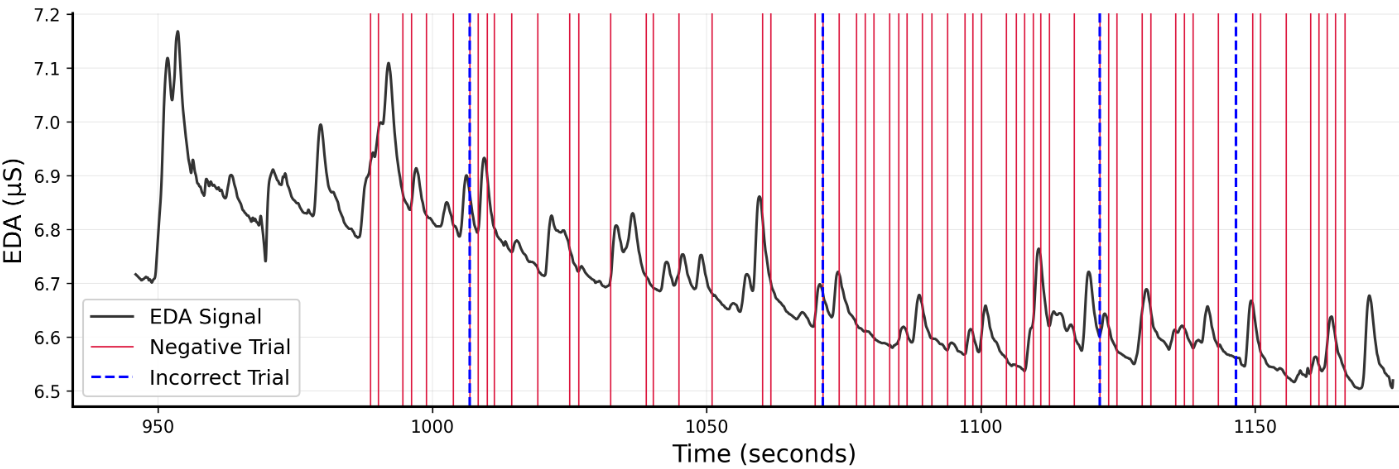}
    \vspace{-0.6cm}
    \caption{Example experiment block from a randomly selected participant from~\cite{precog-methods}, with annotated negative and incorrect trials.}
    \label{fig:header}
    \vspace{-0.5cm}
    
\end{figure}

\begin{itemize}
    \item We design and fit a point process generalized linear model (GLM) for EDA sequences, enabling subject-specific characterization of sympathetic arousal dynamics.
    \item  We show that task-related covariates provide a better fit to the EDA dynamics than individual baseline activity, as demonstrated in an Emotional Stroop task with 131 participants.
    \item We show that point process modeling produces clinically relevant biomarkers of depression and suicidal ideation that improve on the accuracy of established EDA measures.
\end{itemize}

\section{Related Work}

\textbf{Electrodermal Activity as a Depression Marker.}\, Depression has been associated with impairments in cognitive functions such as emotion regulation, attentional control, and decision making~\cite{grahek2019motivation,keller2019paying}. EDA, as an established marker of sympathetic arousal, reflects physiological processes closely tied to these functions~\cite{critchley2002electrodermal}. Altered EDA responses have been observed in individuals with depression, and thus EDA has been proposed as a candidate biomarker for depressive disorder~\cite{kim2019skin}. Importantly, EDA has also shown utility in differentiating individuals with suicidality from those with major depressive disorder~\cite{pruneti2023electrodermal}. A consistent finding across studies is EDA hypoactivity in depressed and suicidal cohorts~\cite{vahey2015galvanic}. Nevertheless, the interpretability of EDA remains constrained by its sensitivity to external influences—such as stimulus timing—which complicates the isolation of disorder-specific or task-specific effects~\cite{alexander2005separating}.

\textbf{Electrodermal Activity as a Point Process.}\, The physiology of SCRs has motivated efforts to disentangle the temporal structure of the recorded pulses~\cite{subramanian2020point}. A variety of approaches have been developed, many of which attempt to model the geometry of SCR waveforms, including their amplitude, rise time, and decay characteristics~\cite{bach2010modelling,benedek2010decomposition}. While potentially informative, such features are often highly idiosyncratic, influenced by individual physiology and recording conditions~\cite{alexander2005separating}. In contrast, our work is inspired by studies that conceptualize SCRs as discrete events and model their emergence and temporal evolution as spike trains~\cite{storm2005palmar}. This event-based perspective has been shown to capture behaviorally relevant dynamics, for example in dyadic interactions~\cite{chaspari2014non, chaspari2015quantifying} and in the assessment of sleep quality~\cite{sano2014quantitative}. Yet, relatively little is known about how point-process representations of EDA under cognitive load might serve as indicators of mental health disruptions such as depression.\vspace{-0.1cm}

\section{The PRECOG Study}

The \textit{Multimodal Integration of Neural and Biobehavioral Signals for Predicting Preconscious Responses} (PRECOG) study investigated the neural and cognitive mechanisms underlying depression and suicidal ideation through a series of two cognitive tasks paired with multimodal neural and physiological recordings~\cite{precog-methods}. A total of 146 college-aged adults completed those tasks and were screened using the Patient Health Questionnaire-9 (PHQ-9) and the Suicide Ideation Scale (SIS). Based on these surveys, they were grouped into three categories: healthy controls (C, 49 subjects), individuals with depression without suicidal ideation (D, 47 subjects), and individuals with depression and suicidal ideation (S, 50 subjects). More details about the study protocol can be found in~\cite{precog-methods}. This paper focuses on the first of the two tasks, namely the \textit{Emotional Stroop}.\vspace{-0.1cm}

\subsection{Emotional Stroop}

During the Emotional Stroop, each participant viewed a total of 480 fast-paced trials on a black screen background. Each trial presented a single word of positive/neutral or negative valence in a colored font. Participants viewed each trial for 0.5 sec and were instructed to identify the font color (red, yellow, green, or blue) as quickly and accurately as possible by pressing a corresponding colored button. Stimuli were presented in randomized order in 4 distinct blocks, with equal block-wise representation across sentiment categories. Sentiment categories comprised words with happy or neutral sentiment (designated as the positive/neutral valence group; 120 words) and sad or suicide sentiment (designated as the negative valence group; 120 words).  Each block started with a resting period of about 30 seconds and continued with 120 word trials, yielding a total duration of about 12 minutes. The task was designed to evoke emotional and cognitive interference markers of depression and suicidality~\cite{precog-methods}.\vspace{-0.1cm}

\subsection{Data Acquisition}

EDA signals were recorded with BrainVision B18 Multitrode electrodes~\footnote{\href{https://shop.easycap.de/products/multitrode}{https://shop.easycap.de/products/multitrode}} at a universal sampling rate of 1~kHz. Two flat electrodes were affixed to the medial phalanges of the index and middle finger of the non-dominant hand. Recordings were subsequently synchronized to trial-level stimulus and response markers. Raw signals were band-pass filtered between 0.67 and 45~Hz using a 5th order, zero-phase Butterworth filter to remove high-frequency noise and powerline interference. Lastly, we identified and excluded 15 participants by visual inspection due to having missing or severely distorted signal acquisition, and were left with 131 subjects for analysis.

\section{Methods}

In this section we describe the pre-processing of EDA recordings, the proposed point process model and its parameterization, as well as the evaluation of the fitted parameters as biomarkers.

\subsection{EDA Event Detection}

EDA is composed of a slow-varying (tonic) component and a number of event-driven spikes (SCRs) that were identified through the cvxEDA decomposition method~\cite{greco2015cvxeda}, which is well-established for this task. The EDA recording of a sample participant that underwent this process is shown in Figure~\ref{fig:header}. We used the Neurokit2 library~\cite{Makowski2021neurokit} to detect SCRs in the phasic component of each signal. Importantly, the detected SCRs were aligned to their \textit{onset} rather than \textit{peak} times for fitting the point process models. SCR onsets preserve the precise temporal relationship between the stimulus and the initiation of sympathetic arousal, whereas peak latencies are more variable and prone to idiosyncratic physiological latency and recording jitter that weaken the association with task covariates.

\subsection{Point Process Model}

We fit a separate point process model per individual. We fit the sequence of SCR events during the Stroop task as a spike train through a non-homogeneous Poisson process, where the conditional intensity captured both a baseline rate and modulations time-locked to Stroop trials. Let $\rho_j$ denote the time of the participant's response on a given trial $j$. The conditional intensity function $\lambda(t)$ governs the probability of observing an SCR event in $[t, t+dt)$:
\begin{equation}
    \lambda(t) = \mu + \sum_{j=1}^N A_j\,e^{-\frac{t - \rho_j}{\tau}}\, \mathbf{1}\{t \ge \rho_j\},
\end{equation}
where $\mu$ is the baseline event rate, $N$ is the total number of trials (here $N=480$), $A_j$ is the amplitude of the kernel associated with trial $j$, and $\tau$ is the decay constant of the kernels. The exponential kernel form models the gradual return of SCR event probability to the baseline following a trigger event~\cite{chaspari2014non}. We defined trial event times based on participants’ response timestamps rather than stimulus onset, under the expected SCR delay and the assumption that the former is a stronger indicator of stimulus receipt. Hence, the indicator function ensures kernels are active only after subject response. 

Covariates were incorporated in the log-intensity domain through a Generalized Linear Model (GLM) to account for systematic variation in EDA responses induced by the Emotional Stroop design. Specifically, the event amplitude parameter $A_j$ was modeled as an exponential function of three task variables:
\begin{equation}
    A_j = A_0\,\exp\!\left(w_{\text{neg}}\,x_{\text{neg}} \,+\, w_{\text{rt}}\,x_{\text{rt}} \,+\, w_{\text{err}}\,x_{\text{err}}\right)
\end{equation}
where $x_{\mathrm{neg}}$ encodes negative-valence trials (1 for sad and suicide-related words, 0 otherwise), $x_{\mathrm{rt}}$ is the z-scored, log-transformed reaction time, and $x_{\mathrm{err}}$ is a binary indicator of incorrect (or missed) responses. This parameterization allows disentangling the baseline electrodermal activity from modulation by word valence, cognitive load, and error processing during trial stimuli; thus three corresponding coefficients $w_{\text{neg}}, w_{\text{rt}}, w_{\text{err}}$ are introduced to quantify the strength and direction of each covariate influence.

\textbf{Physiological translation:}\, In the proposed framework, $\mu$ reflects an individual's non-specific EDA, i.e., independent of task events. The amplitude $A_0$ and $w_{\text{neg}}, w_{\text{rt}}, w_{\text{err}}$ capture the magnitude and modulation of sympathetic responses elicited from the task execution. In this parameterization, $A_0$ holds the strength of EDA responses to the average neutral trial. As for the covariates, positive $w_{\text{neg}}$ means heightened reactivity to negative words, negative $w_{\text{err}}$ suggests blunted responses on errors, whereas $w_{\text{rt}}$ links response vigor to motor-cognitive effort through reaction time. The decay constant $\tau$ is the effective time scale over which elevated SCR probability persists after participant response. By fitting this model at the individual level we obtain a compact, interpretable set of parameters that summarize each participant’s sympathetic activity and arousal profile over the entire Emotional Stroop task.\vspace{-0.2cm}

\subsection{Parameter Estimation}

Given aggregated event counts $y_k$ in time bins of width $dt$, the negative log-likelihood of the model is:
\begin{equation}
    \mathcal{L}(\theta)\,=\,\sum_k \lambda(t_k;\theta)\,dt\,-\,\sum_k y_k\,\log\!\big[\lambda(t_k;\theta)\,dt\big],
\end{equation}
where $\theta = \{\mu, A_0, w_{\text{neg}}, w_{\text{rt}}, w_{\text{err}}, \tau\}$. We optimized $\theta$ for each subject separately using L-BFGS-B~\cite{byrd1995limited}, applying physiologically informed boundaries to ensure plausible rates. Ridge penalties on $w_{\text{neg}}, w_{\text{rt}}, w_{\text{err}}$ prevented overfitting, with stronger regularization on the error term given its sparsity, i.e., less than 5\% of the total trials across the dataset. We set $dt = 1\,$s based on prior work~\cite{chaspari2014non}.

\subsection{Goodness of Fit}

The optimization results were evaluated through log-likelihood and residual analysis metrics. Larger log-likelihood implies a better data fit. To compare the different components of the proposed model, we used Akaike Information Criterion~\cite{akaike1998information} $\text{AIC} = 2P - 2\log L$, where $P$ is the total number of parameters and $L$ is the likelihood value (Equation 3). AIC penalizes the presence of many parameters, with smaller values yielding a better model. Residual analysis was performed with the Kolmogorov-Smirnov (KS) goodness-of-fit test~\cite{massey1951kolmogorov} that compares two Cumulative Distribution Functions (CDFs) $F_1$ and $F_2$ with the statistic $D = \sup x |F_1(x) - F_2(x)|$. Small D indicates that the random samples are likely drawn from the same distribution. Hence, we compare the empirical CDF of real SCR occurrence times to the CDF computed from the proposed model. To quantify model effectiveness, we considered two alternatives:

\textbf{Baseline model.} As a minimal baseline formulation, we modeled SCR events as arising from a homogeneous Poisson process with the following constant intensity $\lambda_1(t) = \mu$, where $\mu$ captures the participant’s idiosyncratic rate of spontaneous SCR events, independent of task engagement. This model can be practically realized as the inverse frequency of SCR events during the task.

\textbf{Baseline trial modulation.} As an intermediate step, we introduced a single kernel triggered at each trial response time $t_j$ with a common amplitude parameter $A_j =A$ across trials (see Equation 1). This model captures the average phasic increase in SCR probability following any Stroop trial, without yet distinguishing between trial-level covariates (i.e., congruency, accuracy, or reaction time). It reflects instead an aggregate trial-locked modulation. 

\textbf{Full model.} In the full specification, we allowed the amplitude $A_j$ of each kernel to vary as a function of trial-specific covariates via a log-linear link as described above. The goal of this experiment was to compare the goodness of fit for the three models to establish the effectiveness of the proposed $\lambda$ (Equation 1).\vspace{-0.2cm}

\subsection{Classification Task}

After fitting the point process model to each participant’s EDA during the Emotional Stroop trials, the resulting parameter vector $\theta = \{\mu, A_0, w_{\text{neg}}, w_{\text{rt}}, w_{\text{err}}, \tau\}$
serves as a subject-level representation of their EDA during the entire task. We use z-scored $\theta$ as the input feature vector to a radial basis function (RBF) support vector machine (SVM) to differentiate depressed and suicidal individuals from healthy controls. We report leave-one-subject-out cross-validation results over the entire set of participants, where no explicit hyperparameter tuning is performed. Instead, we use the default settings provided by the scikit-learn library~\cite{sklearn}. For comparison purposes, we constructed an additional seven-parameter feature vector comprising summary EDA statistics per individual: tonic level mean, variance, and slope; SCR mean amplitude and variance; and SCR mean rise time and variance. Model performance was assessed in terms of area under the receiver-operating curve (AUROC), sensitivity, and specificity with respect to the control class.

\section{Results}

\subsection{Optimization Results}

\begin{table}

\centering
\begin{tabular}{lccc}
\toprule
 & {$\lambda_1$} & {$\lambda_2$} & {$\lambda$} \\
\midrule
\# Parameters & 1 & 3 & 6 \\

NLL $(-\log L)$ & 160.7$_{\pm80.8}$ & 259.2$_{\pm152.9}$ & 156.2$_{\pm83.6}$ \\

AIC & 323.4$_{\pm161.6}$ & 524.4$_{\pm305.8}$ & 324.4$_{\pm167.2}$ \\

KS $\times 10^{-1}$ & 3.4 & 2.8 & 2.4 \\

\bottomrule
\end{tabular}
\caption{Model comparison for EDA point process analysis. NLL refers to negative log-likelihood. KS refers to Kolmogorov-Smirnov statistic. For all metrics, lower scores indicate a better model fit.}
\label{tab:optim}
\vspace{-0.3cm}
\end{table}

The optimization results in Table~\ref{tab:optim} indicate clear differences in fit and complexity across the three parameterizations. The homogeneous model $\lambda_1$ provides a relatively low negative log-likelihood (160.7) but at the cost of higher misfit according to the KS statistic. The three-parameter $\lambda_2$ introduces additional flexibility but does not translate into improved fit; both the NLL and the AIC are substantially worse, reflecting over-parameterization relative to the available information. In contrast, the six-parameter model $\lambda$ achieves the lowest NLL (156.2) and the best KS statistic (0.24), while maintaining an AIC comparable to the most parsimonious configuration. 

Notably, our model $\lambda$ incurs an AIC penalty of 10 points compared to the homogeneous $\lambda_1$ since it has 5 additional parameters, yet the measured fit difference is only a single point.  This suggests that the added expressiveness captures meaningful structure in the point process without introducing excessive variance. Further, given the baseline threshold for goodness-of-fit~\cite{massey1951kolmogorov}, $th = 1.36 / \sqrt{N} = 0.17$, the fit quality can be characterized as moderate to good for the majority of individuals. Overall, these results support the proposed model as the most effective for characterizing SCR event dynamics. Finally, no significant group-level differences in model fit were observed (Kruskal–Wallis test, all $p>0.05$), ruling out systematic bias for subsequent machine learning analyses.\vspace{-0.2cm}

\subsection{Model Parameter Analysis}

The goal of the study is to determine whether the parameters of the individually fit point process models encode markers of depression and suicidality. To identify such group-level differences, we first evaluated the linear discriminability of the six parameters ($\mu$, $A_0$, $w_{neg}$, $w_{rt}$, $w_{err}$, $\tau$) as well as the number of SCRs ($n_{events}$) with a two-sided Mann–Whitney U test for independent groups. No parameter exhibited statistically significant group differences after FDR correction for multiple comparisons ($p>0.05$), except $w_{neg}\, (p=0.031)$. Effect sizes were uniformly small (all Cohen's $d<0.20$), suggesting no systematic shift between groups, which motivated the choice of the non-linear, kernel-based classifier.\vspace{-0.2cm}

\subsection{Classification Results}

\begin{table}
\centering

\begin{tabular}{lccc}
\toprule
\textbf{Metric} & \textbf{Point Process} & \textbf{SCR baseline} & \textbf{Combined} \\
\midrule
\multicolumn{4}{l}{\textit{Control vs. Clinical}}\vspace{0.05cm} \\
AUROC        & \textbf{0.730} $\pm$ 0.049 & 0.571 $\pm$ 0.059 & 0.684 $\pm$ 0.051 \\
Sensitivity  & 0.761 $\pm$ 0.043          & 0.673 $\pm$ 0.049 & \textbf{0.782} $\pm$ 0.044 \\
Specificity  & \textbf{0.692} $\pm$ 0.068 & 0.441 $\pm$ 0.077 & 0.506 $\pm$ 0.074 \\
\midrule
\multicolumn{4}{l}{\textit{vs. Depressed only}}\vspace{0.05cm} \\
AUROC        & \textbf{0.761} $\pm$ 0.052 & 0.598 $\pm$ 0.061 & 0.710 $\pm$ 0.054 \\
Sensitivity  & \textbf{0.834} $\pm$ 0.049 & 0.762 $\pm$ 0.067 & 0.809 $\pm$ 0.061 \\
\midrule
\multicolumn{4}{l}{\textit{vs. Suicidal only}}\vspace{0.05cm} \\
AUROC        & \textbf{0.702} $\pm$ 0.057 & 0.547 $\pm$ 0.064 & 0.661 $\pm$ 0.058 \\
Sensitivity  & 0.694 $\pm$ 0.067          & 0.591 $\pm$ 0.072 & \textbf{0.758} $\pm$ 0.061 \\
\bottomrule
\end{tabular}
\caption{Classification performance under LOSO cross-validation, repeated over 10 random seeds. AUROC and sensitivity are broken down by clinical group. Results shown as mean $\pm$ std.}
\label{tab:classification_results}
\vspace{-0.3cm}
\end{table}

Across all comparisons, the point-process model consistently outperformed the summary SCR statistics. For the \textit{control vs. clinical} setting, the point process achieved the highest AUROC ($0.730 \pm 0.049$) and specificity ($0.692 \pm 0.068$), indicating superior overall discriminability and a lower false-positive rate compared to SCR features alone. The combined feature set did increase sensitivity ($0.782 \pm 0.044$), suggesting that the inclusion of summary SCR measures biased the classifier toward clinical cases. SHAP analysis ranked word valence $w_{neg}$ as the most important feature (Figure~\ref{fig:analysis}-left), with baseline rates $\mu$ and $A_{0}$ showing limited contributions.

These results indicate that the point-process model improves upon the SCR baseline mainly with respect to specificity (0.441 to 0.692), whereas SCR statistics contribute more toward identifying high-risk individuals at the expense of false alarms. Taken together, our findings highlight that temporal modeling of electrodermal dynamics provides a more balanced signal for discriminating control participants from clinical populations. We attribute this improvement to the achieved disentanglement of the fast trial dynamics as well as the avoidance of idiosyncratic elements.\vspace{-0.3cm}

\subsection{Ablation on Covariates}

To evaluate the contribution of each fitted parameter and covariate to the classification performance, we conducted an ablation analysis in which features were systematically removed one at a time. A classifier was re-trained with each feature excluded in turn.

\begin{figure}
    \centering
    \includegraphics[width=\linewidth]{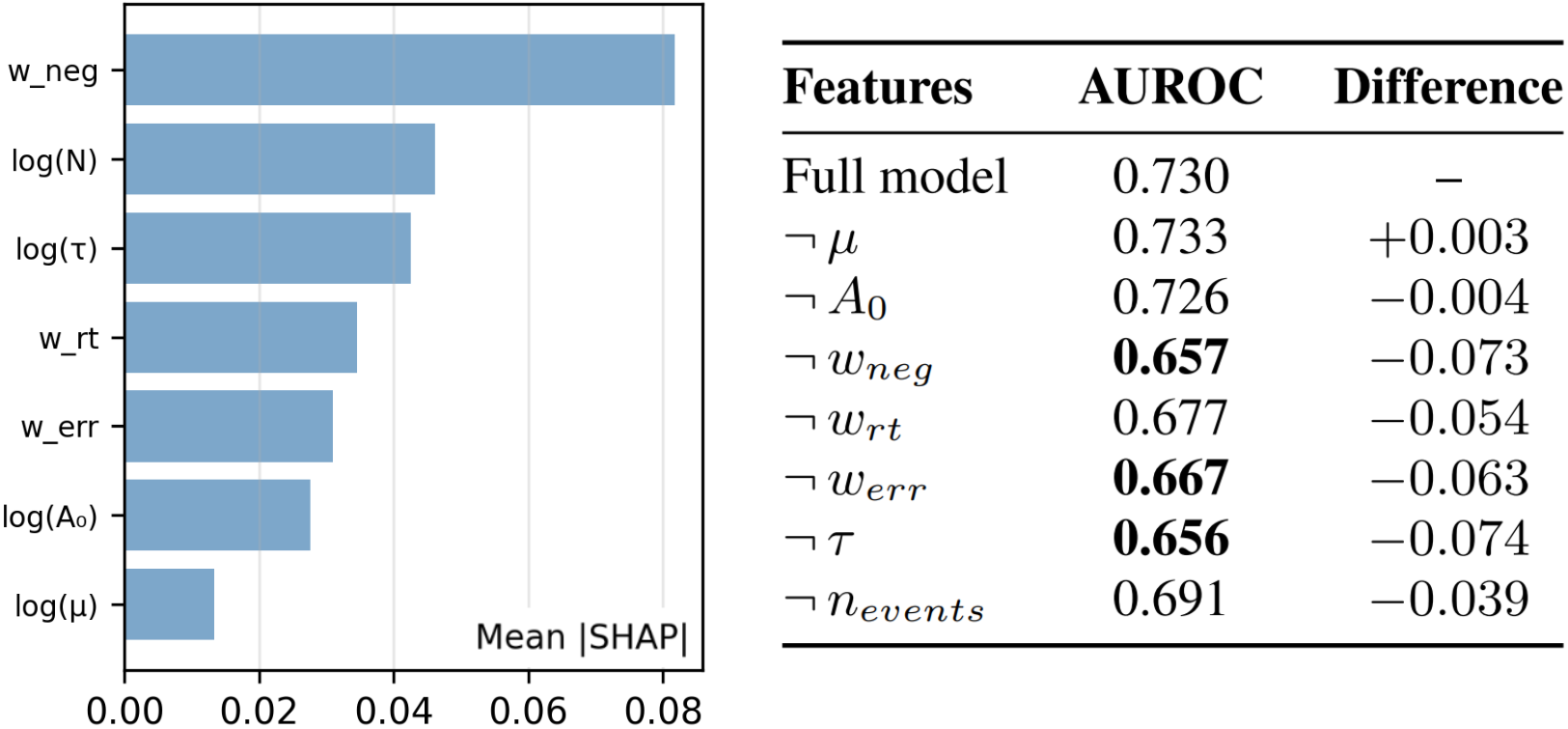}
    \vspace{-0.4cm}
    \caption{Left: Feature importance values per SHAP analysis. Right: Ablation analysis of covariates. AUROC is reported for each feature removed. All variances were similar ($\approx 0.05$). Bold values denote settings where the performance drop was statistically significant.}
    \vspace{-0.4cm}
    \label{fig:analysis}
\end{figure}

As shown in Figure~\ref{fig:analysis}-right, removing $\mu$ or $A_{0}$ had negligible effect on performance ($+0.003$ and $-0.004$, respectively), hence these parameters provided limited unique predictive value beyond the remaining features. In contrast, omitting dynamic weighting terms such as $w_{neg}$, $w_{rt}$, or $w_{err}$ resulted in notable decreases in AUROC ($-0.073$, $-0.054$, and $-0.063$, respectively). Likewise, exclusion of $\tau$ produced the largest regression in performance ($-0.074$), underscoring its importance for capturing temporal dynamics of the process. Finally, removing $n_{events}$ reduced AUROC by $-0.039$, implying a moderate contribution. Overall, our analysis shows that parameters encoding temporal and error-related dynamics ($w_{neg}$, $w_{rt}$, $w_{err}$, and $\tau$) were critical to discriminative performance, whereas baseline rates ($\mu$, $A_{0}$) played a minimal role. These observations are aligned with our findings from both the goodness-of-fit analysis and the feature importance analysis.\vspace{-0.1cm}

\section{Discussion}
\vspace{-0.1cm}
Our findings demonstrate that a Poisson point process model of electrodermal activity provides a physiologically intuitive framework for capturing short-term sympathetic responses. SCRs arise from the stochastic bursting of sudomotor nerve activity, where discrete neural discharges shape overlapping phasic waveforms. Conventional statistics average over these bursts, thereby discarding information about their precise timing and dynamics. In contrast, our point process formulation explicitly models SCRs as events generated by an underlying stochastic intensity function. This mechanistic alignment with sympathetic physiology could explain the superior performance across all clinical comparisons, relative to SCR statistics.

Although none of the fitted parameters exhibited large group-level differences, the ablation analysis and feature-level inspection revealed a consistent signal: baseline parameters ($\mu$, $A_0$) contributed little to classification, whereas temporal weighting terms, particularly related to valence $w_{neg}$, accounted for the largest decrements in performance when removed. This was also observed in the goodness of fit analysis, for which models lacking expressive temporal structure showed poorer calibration. From a clinical perspective, the identification of $w_{neg}$ as a key discriminative parameter highlights altered reactivity to negative stimuli as a potential physiological correlate of depression and suicidality. These findings align with prior psychophysiological evidence linking negative-affect processing to autonomic dysregulation. Further, the model tended to identify depressed individuals better than suicidal ones, which could imply a suicide-related moderator factor in EDA responses.

That said, several limitations should be acknowledged. First, the modest sample size constrains the generalizability of our findings. Second, while the Stroop paradigm provides a controlled stressor, it does not capture the full range of stressors encountered in daily experiences; model performance should therefore not be interpreted as reflecting the limits of predictability. Third, the models were trained on relatively short experimental sessions. Although this design helped mitigate confounds such as participant fatigue, it could not reflect long-term stress dynamics. Future work should scale to larger and more diverse populations and incorporate complementary neural and behavioral measures to improve precision in risk assessment

\vfill\newpage
\bibliographystyle{IEEEbib}
\bibliography{refs}

\end{document}